\documentstyle[12pt,epsf]{article}
\begin{document}
\title{Monopoles and dyons in $SO(3)$ gauged Skyrme models}

\author{{\large Y. Brihaye}$^{\star}$,
{\large B. Hartmann}$^{\diamond}$
and {\large D. H. Tchrakian}$^{\dagger}$ \\ \\
$^{\star}${\small Physique-Mathematique, Universite de Mons-
Hainaut, Mons, Belgium}\\ \\
$^{\diamond}${\small Fachbereich Physik, Universit\"at Oldenburg,
Postfach 2503, D-26111 Oldenburg, Germany}\\ \\
$^{\dagger}${\small Department of
Mathematical Physics, National University of Ireland Maynooth,} \\
{\small Maynooth, Ireland} \\ {\small and} \\
{\small School of Theoretical Physics -- DIAS, 10 Burlington Road,
Dublin 4, Ireland }}

\date{}
\newcommand{\dd}{\mbox{d}}\newcommand{\tr}{\mbox{tr}}
\newcommand{\ee}{\end{equation}}
\newcommand{\eea}{\end{eqnarray}}
\newcommand{\be}{\begin{equation}}
\newcommand{\bea}{\begin{eqnarray}}
\newcommand{\ii}{\mbox{i}}\newcommand{\e}{\mbox{e}}
\newcommand{\pa}{\partial}\newcommand{\Om}{\Omega}
\newcommand{\vep}{\varepsilon}
\newcommand{\bfph}{{\bf \phi}}
\newcommand{\lm}{\lambda}
\def\theequation{\arabic{equation}}
\renewcommand{\thefootnote}{\fnsymbol{footnote}}
\newcommand{\re}[1]{(\ref{#1})}
\newcommand{\bfR}{{\sf R\hspace*{-0.9ex}\rule{0.15ex}%
{1.5ex}\hspace*{0.9ex}}}
\newcommand{\N}{{\sf N\hspace*{-1.0ex}\rule{0.15ex}%
{1.3ex}\hspace*{1.0ex}}}
\newcommand{\Q}{{\sf Q\hspace*{-1.1ex}\rule{0.15ex}%
{1.5ex}\hspace*{1.1ex}}}
\newcommand{\C}{{\sf C\hspace*{-0.9ex}\rule{0.15ex}%
{1.3ex}\hspace*{0.9ex}}}
\newcommand{\eins}{1\hspace{-0.56ex}{\rm I}}
\renewcommand{\thefootnote}{\arabic{footnote}}

\maketitle
\begin{abstract}
Three dimensional $SO(3)$ gauged Skyrme models characterised by
specific potentials imposing special asymptotic values on the
chiral field are considered. These models are shown to support finite
energy solutions with nonvanishing magnetic and electrix flux, whose
energies are bounded from below by two distinct charges -- the magnetic
(monopole) charge and a non-integer version of the Baryon charge. Unit
magnetic charge solutions are constructed numerically and their properties
characterised by the chosen asymptotics and the Skyrme coupling are
studied. For a particular value of the chosen asymptotics, charge--2
axially symmetric solutions are also constructed and the attractive
nature of the like--monopoles of this system are exhibited. As an
indication towards the possible existence of large clumps of monopoles,
some consideration is given to axially symmetric monopoles of
charges-2,3,4.
\end{abstract}
\medskip
\medskip
\newpage

\section{Introduction}

In the $O(4)$, or usual, Skyrme~\cite{S} model in 3 dimensions, the
finite energy conditions specify the asymptotic value at large distances
of the chiral field uniquely. Depending on the parametrisation, either
the $SU(2)$ valued field $U$ or the $S^3$ valued field $\phi^a$, subject
to $|\phi^a|^2$, with $a=1,2,3,4$, this asymptotic value is
\be
\label{asym}
\lim_{r\to \infty}U=\eins\ \quad {\rm or}\quad
\lim_{r\to \infty}\phi^4=1\ .
\ee
The fields $U$ and $\phi^a$ are related through
\be
\label{u}
U=\phi^a\ \sigma^a\ ,\qquad U^{\dagger}=\phi^a\ \tilde \sigma^a\ ,
\ee
where in terms of the Pauli matrices $\vec \tau$,
$\sigma^a=(i\tau^i ,\eins)$ and $\tilde\sigma^a=(-i\tau^i ,\eins)$.

Often, the static Hamiltonian of the Skyrme system is augmented with a
'pion-mass' potential
\be
\label{pimass}
V_{\pi}(\phi^a)=m_{\pi}(1-\phi^4)\ ,
\ee
consistent with the asymptotics \re{asym}. The only practical effect that the inclusion of this potential \re{pimass} has is, that it renders the
asymptotic behaviour of the chiral function exponential, where in its
absence this would have been a power decay.

The situation is very different when the Skyrme model is
gauged in one~\cite{gauged} or other~\cite{BT} gauging prescription,
as a result of which the asymptotic value of
the chiral field is not fixed uniquely by finite energy conditions.
This feature of 3 dimensional gauged Skyrme models was considered and
highlighted in \cite{KTZ}.

In the present work, we augment the 3 dimensional $SO(3)$ gauged
Skyrme model studied in \cite{AT,BT}, with the potential\footnote{Like
\re{pimass}, this potential is also chosen such that it results in
the exponential behaviour of $\phi^4$ asymptotically.}
\be
\label{pot}
V=\lambda(\cos \omega -\phi^4)^2\ ,\qquad \pi \ge \omega \ge 0\ ,
\ee
whose effect is to specify the asymptotic values of the chiral field
uniquely, consistent with finiteness of the energy. The new asymptotics
are
\be
\label{chirasym}
\lim_{r\to 0}\phi^4=-1\, \qquad \lim_{r\to \infty}\phi^4=\cos \omega\ .
\ee

It is clear from \re{chirasym} that the volume integral of the density
that maps the field space to the configuration space, is not going to be
an integer except in the case where $\omega=0$. Thus the lower bounds
labeled by this charge cannot be identified with the degree of the map,
or the topological Baryon charge, except when $\omega=0$. Such a
noniteger charge however does supply a legitimate lower bound on the
energy integral. For want of a better name, we shall persist in calling
such charges $Q_B(\omega)$, with the understanding that only $Q_B(0)$
is really the Baryon charge.

In the generic case $\pi \ge \omega \ge 0$, there will be an independent
lower bound in addition to $Q_B(\omega)$, namely the magnetic monopole
flux $\mu(\omega)$, which also depends on $\omega$. These lower bounds
will be stated explicitly below. The main feature of the dynamics
characterised by the potential \re{pot}, with $\omega\neq0$ is that the
$SO(3)$ is broken down to $U(1)$, with the residual Maxwell field
described by the correponding '~Hooft--tensor supporting a magnetic flux.
Like the charge $Q_B(\omega)$, this magnetic flux $\mu(\omega)$ is integer
also only modulo a continuous factor depending on $\omega$, and
takes an integer value only when $\omega=\frac{\pi}{2}$. The solutions
we have found turn out, as expected, to respect both lower bounds
$\mu(\omega)$ and $Q_B(\omega)$.

We have confirmed the existence of finite energy solutions bounded by
the charges $\mu(\omega)$ and $Q_B(\omega)$ by numerical construction.
An interesting result is that when $\omega \neq 0$, the solution persists
even when the Skyrme coupling constant vanishes, i.e. $\kappa_2=0$ in
\re{L} and \re{H}. This is not surprising since the presence of the
Yang--Mills term satisfyies the (Derrick) scaling requirement independently
of the Skyrme term, and in this case, the soliton is bounded from below
only by the magnetic flux $\mu$. This will be explained in more detail
in Section {\bf 2}, where the model is defined and the two said lower
bounds will be stated. Then in Section {\bf 3}, we study the spherically
symmetric solutions for various $\omega$ and $\kappa$, and find the
ranges of these two parameters for which solutions exist by numerical
construction. In Section {\bf 4}, we study the axially symmetric
magnetic charge--2 solution for the particular value of the parameter
$\omega=\frac{\pi}{2}$, with a view to learning whether two like monopoles
of that system can be in an attractive or a repulsive phase. We encounter
the rather surprising result that even for $\kappa_2=0$ this is
{\em attractive}, and then as expected it becomes even more attractive
with increasing $\kappa_2>0$. Section {\bf 5} is devoted to summarising
and discussing our results.

\section{The model and lower bounds}

The model is specified by the gauging prescription and is the 3
dimensional model used in \cite{AT,BT}, augmented by the potential
\re{pot}. Usually, we will treat this potential as a {\it gedanken}
entity and will not exploit it save as an agency justifying the
asymptotics \re{chirasym}. In terms of the $S^3$ valued field
$\phi^a=(\phi^{\alpha},\phi^4)$, $\alpha=1,2,3$, and the $SO(3)$ gauge
connection $A_{\mu}^{\alpha}$ with curvature $F_{\mu \nu}^{\alpha}$,
the covariant derivative is defined by the prescription
\be
\label{cov}
D_{\mu} \phi^{\alpha} =\pa_{\mu} \phi^{\alpha} 
+\vep^{\alpha \beta \gamma}A_{\mu}^{\beta}\
\phi^{\gamma},
\qquad \qquad D_{\mu}\phi^{4} =\pa_{\mu} \phi^{4}\ .
\ee

Since much of the analysis will be almost
identical to that in (the relevant) Section {\bf 3} of \cite{BT}, we
will use the same notation here. This will enable us to present some
of the new results without the necessity of repeating the detailed
analyses leading to them. The model is described by the Lagrangian
\be
\label{L}
{\cal L}=
-\kappa_0^4 |F_{\mu \nu}^{\alpha}|^2+{1\over 2}
\kappa_1^2 |D_{\mu}\phi^a|^2 -{1\over2}\kappa_2^4
|D_{[\mu}\phi^a D_{\nu]}\phi^b|^2-V(\phi^4) \,
\ee
which in the temporal gauge $A_0^{\alpha}=0$ yields the static Hamiltonian
\be
\label{H}
{\cal H}=\kappa_0^4 |F_{ij}^{\alpha}|^2+{1\over 2}
\kappa_1^2 |D_{i}\phi^a|^2 +{1\over
2}\kappa_2^4 |D_{[i}\phi^a D_{j]}\phi^b|^2+V(\phi^4) \ .
\ee
The potential $V(\phi^4)$ in both \re{L} and \re{H} is that given by
\re{pot}. In our study of the 'monopole', we will be mainly concerned with
the energy density functional \re{H}, but we give the corresponding
Lagrangian \re{L} too in anticipation of our discussion of the
corresponding 'dyon' solution.

We proceed to state the two distinct lower bounds on the energy, namely
the volume integral of \re{H}. Both bounds, the 'magnetic monopole' charge
and the noninteger 'baryon charge', pertain to the generic asymptotics
\re{chirasym} with $\omega\neq0$, dictated by \re{pot}.

The first of these follows from the classic Bogomol'nyi inequality
\be
\label{bog}
\kappa_0^4 |F_{ij}^{\alpha}|^2
+{1\over 2}\kappa_1^2 |D_{i}\phi^{\alpha}|^2
\ge\kappa_0^2\kappa_1\vep_{ijk}\pa_k(\phi^{\alpha}F_{ij}^{\alpha})\ .
\ee
It is obvious that the left hand side of \re{bog} can be replaced by
${\cal H}$ of \re{H}, by adding suitable positive definite terms to it,
resulting in
\be
\label{bogo}
{\cal H}
\ge\kappa_0^2\kappa_1\vep_{ijk}\pa_k(\phi^{\alpha}F_{ij}^{\alpha})\ ,
\ee
on the right hand side of which we recognise the $U(1)$ 't~Hooft-tensor,
$\phi^{\alpha}F_{ij}^{\alpha}$, of the residual gauge field responsible
for the magnetic flux provided that $\omega\neq0$, so that
$|\phi^{\alpha}|\to\sin\omega\neq0$ asymptotically.

To state the corresponding inequality for the other lower bound, we
define the 'baryon charge' density and its covariantised version,
respectively,
\bea
\varrho_0&=&\frac{1}{12\pi^2} \vep_{ijk} \vep^{abcd} \pa_{i}\phi^a
\pa_{j}\phi^b \pa_{k}\phi^c \phi^d \label{varrho0} \\
\varrho_G&=&\frac{1}{12\pi^2} \vep_{ijk} \vep^{abcd} D_{i}\phi^a
D_{j}\phi^bD_{k}\phi^c \phi^d \label{varrhoG}\ .
\eea
The volume integral of \re{varrho0} is the noninteger 'baryon charge'
\be
\label{bar}
\int d^3 x \varrho_0 =(\pi-\omega)N \ ,
\ee
except when $\omega=0$, when it is simply the integer $N$, the degree of
the map or the usual baryon charge. The actual gauge invariant charge
density which enters the relevant inequality is defined in terms of
$\varrho_G$ in \re{varrhoG} by~\cite{AT,BT}
\be
\label{varrho}
\varrho = \varrho_G+3\vep_{ijk}\
\phi^{\alpha} F_{ij}^{\alpha}\ D_k\phi^4\ ,
\ee
whose volume integral turns out to be equal to the 'baryon charge'
\re{bar}.

As shown in \cite{BT}, it follows that
\be
\label{ineq}
{\cal H} \geq 
{\kappa_1 \kappa_2^2 \over \sqrt{1 + 9 ({\kappa_2 \over \kappa_0})^4}}
\ \varrho \ .
\ee

We can now conclude from \re{bog} and \re{ineq} the two distinct lower
bounds on the energy $E=\int d^3 x {\cal H}$, namely the 'magnetic'
and 'baryonic' lower bounds, following from the asymptotics \re{chirasym}
\bea
E&\ge&4\pi\kappa_0^2\kappa_1\sin\omega \label{mono}\\
E&\ge&{12\pi\kappa_1 \kappa_2^2 \over \sqrt{1 + 9
({\kappa_2 \over \kappa_0})^4}}(\pi-\omega)\label{baryo}\ .
\eea

The two inequalities \re{mono} and \re{baryo} signal the possibility of
finding finite energy solutions bounded from below, provided that the
(Derrick) scaling requirement is satisfied, which for \re{H} in 3
dimensions, it is. For the limiting case of $\omega=0$ considered in
\cite{BT}, inequality \re{mono} trivialises and \re{baryo} then coincides
with the lower bound used in \cite{BT}. In the other limit when
$\omega=\pi$, both \re{mono} and \re{baryo} trivialise so we would
expect to find no nontrivial solutions in this case. This will be
confirmed by our numerical results to be given below. For
generic values of $\omega$ between these two limiting values, both lower
bounds are valid independently, and any nontrivial finite energy solution
must respect these. This will also be confirmed by our numerical results.

It should perhaps be pointed out that neither of the bounds \re{bogo}
and \re{ineq} can be saturated. As we shall see in Section {\bf 3}, for
$\omega\neq 0$ finite energy solutions persist also for $\kappa_2=0$. But
even in that case, the inequality \re{bogo} cannot be saturated. Thus
in the $\kappa_2=0$ model, we have a system which does not saturate a
Bogomol'nyi bound and whose stress-energy tensor therefore never vanishes.
It follows that the charge-2 monopole of this model is {\em either}
attractive {\em or} repulsive, a property which it shares with the usual
(ungauged) Skyrme model~\cite{S}, in the latter case as is well known
it being attractive~\cite{KS}. We shall find in Section {\bf 4} that the
model with $\kappa_2=0$ supports solutions describing mutually attracting
like monopoles. Then as expected, when the Skyrme coupling constant
$\kappa_2$ is switched on this binding energy will grow further as is
usual with other theories~\cite{KOT} involving Skyrme like kinetic terms.

\section{Spherically symmetric solutions}
This Section is divided into two Subsections, in the first of which we
present the reduced one dimensional subsystems of \re{L} and \re{H},
while in the second we present our numerical results.

\subsection{One dimensional subsystems}
As in \cite{BT}, we impose the spherical symmetry thus
\be
\label{spha}
A_0^{\alpha}=\kappa_1^{-1}g(r)\ \hat x^{\alpha}\ , \qquad
A_i^{\alpha}=\frac{a(r)-1}{r}\ \vep_{i\alpha\beta}\ \hat x^{\beta}\ ,
\ee
\be
\label{sphf}
\phi^{\alpha}=\sin f(r)\ \hat x^{\alpha}\ ,\qquad \phi^4=\cos f(r)\ .
\ee

The ensuing reduced one dimensional Lagrange and (static) Hamiltonian,
subject to a suitable rescaling $r\to x$ such that all constants but
the Skyrme coupling $\kappa_2^4\equiv \kappa$ are suppressed, are
respectively
\bea
\label{rL}
L&=&-2\left(2a'^2+\frac{(a^2-1)^2}{x^2}\right)
-\frac{1}{2}(x^2f'^2+2a^2\sin^2f)
-2\kappa a^2\sin^2f\left(2f'^2+\frac{a^2\sin^2f}{x^2}\right)
\nonumber \\
&&+x^2g'^2\ +\ 2a^2 g^2 \label{rL}
\eea
\be
\label{rH}
H=2\left(2a'^2+\frac{(a^2-1)^2}{x^2}\right)
+\frac{1}{2}(x^2f'^2+2a^2\sin^2f)
+2\kappa a^2\sin^2f\left(2f'^2+\frac{a^2\sin^2f}{x^2}\right)
\ee
Note that we have suppressed the potential terms arising from \re{pot}
in \re{rL} and \re{rH}, since as will be explained below, nearly all
our numerical constructions will be carried out in the $\lambda=0$ limit.

While \re{rH} is positive definite, \re{rL} is not. The latter will be
relevant only in some remarks below concerning the dyon solution of
this system.

Let us first consider the case of main interest, namely the monopole
solutions of the equations following from the static energy density
functional \re{rH}.
The finite energy conditions require the following asymptotic values
\be
\label{asymf}
\lim_{x\to 0}f(x)=\pi\ , \qquad \lim_{x\to \infty}f(x)=\omega\ 
\ee
\be
\label{asyma}
\lim_{x\to 0}a(x)=1\ , \qquad \lim_{x\to \infty}a(x)=0\ ,
\ee
as long as $\omega\neq 0$\ . This is the case of interest in the present
work. (The $\omega= 0$, which is only a limiting case here, was studied in
detail in \cite{BT}.)

The behaviours of the functions $f(x)$ and $a(x)$ in the $x\ll 1$ region
are independent of the value of $\lambda$ in \re{pot}, and they are
\bea
f(x)&=&\pi +F_1 x +o(x^3)\ , \quad (x\ll 1) \label{llf} \\
a(x)&=&1+A_1 x^2 +o(x^4)  , \quad (x\ll 1)\label{llf}
\eea

In the $x\gg 1$ region, the asymptotic behaviour of the function $a(x)$
is again independent of the value of $\lambda$ and is
\be
\label{gga}
a(x)=A\ e^{-\frac{1}{2}x \sin \omega }\ ,\quad A={\rm const.}\ ,
\ee
while that of the function $f(x)$ does depend on $\lambda$.
In the limit of vanishing $\lambda$ and finite $\lambda$, these are
respectively the power and exponential decays
\bea
\label{ggf}
f(x)&=&\omega + \frac{F}{x}+o\left(\frac{1}{x^2}\right)\ ,
\label{ggf0} \\
f(x)&=&\omega + \tilde F\frac{e^{-\sqrt{2\lambda}x}}{x}\label{ggflam}\ ,
\eea
where $F$ and $\tilde F$ are constants which can be evaluated by the
numerical process.

Thus, like with the usual (ungauged) Skyrme model, the addition of
this potential results in the exponential localisation of the chiral
function $f(r)$. In the numerical work presented in the next two
Subsections we have used $\lambda=0$ throughout, since the qualitative
properties of the solutions are unchanged when $\lambda>0$. This was
verified in many typical cases and thereafter the potential \re{pot}
played a {\it Gedanken} r\^ole, rather like for the Prasad--Sommerfield
limit of the Georgi--Glashow model, except that here the $\lambda=0$
limit is not particularly interesting since it does not lead to the
saturation of any Bogomol'nyi bound.

\subsubsection{Dyon solution}
Before proceeding to describe our numerical results concerning the
monopole solutions of this model, we briefly allude to the corresponding
dyon solutions. Following Julia and Zee~\cite{JZ} we vary the energy
density \re{rL} with respect to the functions $a(r)$, $g(r)$ and $f(r)$.
Since \re{rL} is not positive definite, the radial Ansatz \re{spha} and
\re{sphf} is not guaranteed to be consistent with the full
Euler--Lagrange equations of this system. It has however been verified
in \cite{BKT} that \re{spha}--\re{sphf} is indeed consistent.  

The crucial equation that signals the existence of a dyon solution is
the $r\gg1$ asymptotic equation arising from the variation of the
function $a(x)$
\be
\label{dyon}
a''=a\left(\frac{a^2-1}{x^2}-g^2+\sin^2f+...\right)
\ee
which in the $x\to\infty$ limit reduces to
\[
a''=(\sin^2\omega-g^2)a\ ,
\]
which yields acceptable exponentially decaying solutions only when
the asymptotic value of $q=\lim_{x\to\infty}g(x)$ satisfies the condition
\be
\label{condition}
0\le q\ \le\ \sin\omega\ ,
\ee
and otherwise leading to unacceptable oscillatory behaviour for the
function $a(r)$ asymptotically.

With the asymptotic condition \re{condition}, one has the behaviour
\be
\label{elec}
g(x)=q-\frac{c}{x}+o(x^{-2})\ ,
\ee
in which the constant $c$ is evaluated by the numerical integrations and
parametrises the electric flux of the dyon. We do not repeat here the
detailed results of of the numerical process as this is identical to that
for the dyon~\cite{JZ} of the Georgi--Glashow model, as presented in
\cite{BKT}. The relevant analysis in Section {\bf V} of Ref.\cite{BKT}
can be adapted to the present model, by substituting the condition
$0\le q\le 1$ there, by \re{condition}.

The only qualitative difference between the dyon of the Georgi-Glashow
model and the dyon of the present model is, that unlike in the former
case here there is no Prasad--Sommerfield limit.

\subsection{Numerical results}

Solving the spherically symmetric equations for numerous values of the
parameters $\kappa \in ]0,\infty[$ and $\omega \in ]0,\pi[$, strongly
indicates that they admit at least one regular, finite-energy solution for
each choice of these parameters.

The behaviour of the solution in the limit $\omega \rightarrow 0$ is
different according to the value of $\kappa$ and is strongly influenced by
the pattern of solutions occuring in the case $\omega=0$.  We briefly
recall (see \cite{BT,BKT} for further details) that
for $\omega = 0$, solutions with $a(\infty)=0$ exist only for
$\kappa > \kappa_{cr}$ , $\kappa_{cr}\approx 0.697$. 
For $\kappa \in ]0.0,0.697[$ the relevant solution has $a(\infty) = 1$.

Let us now describe how the solutions look like for fixed $\kappa$, with
$\omega$ varying. We have found it convenient to characterise the
solutions by the value of the asymptotic coefficient $F$ defined in Eq.
(27).
The evolution of this parameter is reported in Fig. 1 for several
values of $\kappa$.
For $\kappa > \kappa_{cr}$ the solutions are such that the function
$f(x)$ (resp.
$a(x)$)  decreases monotonically from $\pi$ (resp. $1$) for $x=0$ to
$\omega$ (resp. $0$) for $x\to\infty$. 
The parameter $F$ is positive as seen in Fig. 1 for $\kappa = 1.0.$
The classical energy decreases
monotonically when $\omega$ increases. In the limit $\omega \rightarrow 0$
the classical solution of \cite{BT} are smoothly approached.

The behaviour of the solutions is more elaborate when
 $\kappa < \kappa_{cr}$;
this is illustrated on Fig. 1 and on Fig. 2 for $\kappa = 0.1$ . 
One new feature is that the parameter $F$ undergoes
a change of sign when $\omega$ varies from $0$ to $\pi$.
For large enough values of $\omega$
(say, $\omega > \omega_{max}$, $\omega_{max} \approx 0.5$
 in the case $\kappa=0.1$),
the profiles of $f(x)$ and of $a(x)$ 
monotonically decrease as functions of $x$ and the classical energy 
decreases for $\omega$ increasing.  
For $\omega < \omega_{max}$  the function $f(x)$ develops
a local minimum at an intermediate value of $x$ and the parameter
$F$ defined in Eq.(27) becomes negative.
Moreover, the function $a(x)$ develops a local minimum and a local
maximum at finite values (say $x_m$, $x_M$) of the radial variable.
In this region of $\omega$ the classical energy increases with $\omega$.
When the limit $\omega \rightarrow 0$ is considered
our numerical analysis indicates that $x_m$ stays finite, $x_M$
increases and the value $a(x_M)$ approaches $a=1$ in such a way 
that the
corresponding profile of the $\omega=0$ solution is approached on 
$[0,x_M]$. The corresponding value of the classical energy is also
reproduced.
These different features are illustrated on Figs. 1,2.

Considered as a function of $\omega$, we also observed
(see Fig. 2) that
the classical energy is maximal at an intermediate value 
of $\omega$.
Our numerical analysis indicates that the maximum is
attained precisely for $\omega=\omega_{max}$, i.e. at the value where
the change of sign of the parameter $F$ occurs.
The $\kappa$-dependence of $\omega_{max}$ is reported
on Fig. 3.
 
As further suggested by Fig. 1,
in the region $\kappa \approx 0.5$ , $\omega \approx 0.05$ 
the pattern of the solutions becomes  very complicated.
We obtained strong numerical evidence that several branches
of solutions exist in this region. That is to say e.g, that we find
more than one solution for $\kappa= 0.5$, $\omega=0.05$.
However these new branches seem to exist
on a very small domain of  the
parameter $\omega$,  the numerical analysis
is therefore rather difficult in this region.
Since the study of such details is not the aim of the present paper,
we refrained from further pursuing our numerical analysis in this region.

To finish this section, we mention that, choosing $\kappa=0$,
we were able to construct numerical solutions for $\omega > 1.1729$.
The analysis of the  solutions in the limit $\kappa \rightarrow 0$
(with fixed $\omega < 1.1729$), seems to lead to
a discontinuity of the function $f(x)$.

\section{Axially symmetric solutions}

This Section is divided in two Subsections as in the previous Section,
in the first of which the axially symmetric Ansatz, and, the boundary
conditions of the axially symmetric solutions, are stated. In the second
Subsection, the numerical results are given.

Our objective in this Section is to construct higher magnetic charge
solutions, and in the first instance charge-2 axially symmetric solutions,
with the aim of discovering whether like monopoles of this model are in
an attractive or a repulsive phase. In this framework, we will restrict
our analysis to monopole rather than dyon solutions, in the temporal
gauge $A_0=0$.

The analysis carried out in this Section is less general than that given
in the previous Section for the spherically symmetric solutions. There,
we studied the detailed dependence of the solutions on the parameter
$\omega$ specifying the dynamics. Having exposed these properties
satisfactorily, we proceed to study the most natural subset of models
here, namely those specified by $\omega=\frac{\pi}{2}$, supporting
monopoles of {\it integer} magnetic charges.

Within this $\omega=\frac{\pi}{2}$ subset of models, we consider the
models specified by the Skyrme coupling $\kappa_2$, or, the effective
parameter $\kappa$ for the range $\kappa\ge 0$ which includes
interestingly the point $\kappa=0$.

\subsection{Ansatz and boundary conditions}

With magnetic charge, or azimuthal winding, $n=1,2,3,..$,
the axially symmetric Ansatz~\cite{KKT} for the gauge field is 
\begin{equation}
A_\mu dx^\mu =
\frac{1}{2r} \left[ \tau^n_\phi 
 \left( H_1 dr + \left(1-H_2\right) r d\theta \right)
 -n \left( \tau^n_r H_3 + \tau^n_\theta \left(1-H_4\right) \right)
  r \sin \theta d\phi \right]
\ , \label{gf1}
\end{equation}
and for the Skyrme field it is
\begin{equation}
U=\frac{1}{2}\left( \cos f\ \eins +\sin f \left[(\sin g \sin
\theta + \cos g \cos\theta)\tau_r^{n}+(\sin g \cos\theta -\cos g
\sin\theta)\tau_\theta^{n}\right]\right)
\ . \end{equation}
where $H_1, H_2, H_3, H_4, f$ and $g$ are functions of the coordinates $r$ and
$\theta$.
The symbols $\tau^n_r$, $\tau^n_\theta$ and $\tau^n_\phi$
denote the dot products of the cartesian vector
of Pauli matrices, $\vec \tau = ( \tau_x, \tau_y, \tau_z) $,
with the spatial unit vectors
\begin{eqnarray}
\vec e_r^{\, n}      &=& 
(\sin \theta \cos n \phi, \sin \theta \sin n \phi, \cos \theta)
\ , \nonumber \\
\vec e_\theta^{\, n} &=& 
(\cos \theta \cos n \phi, \cos \theta \sin n \phi,-\sin \theta)
\ , \nonumber \\
\vec e_\phi^{\, n}   &=& (-\sin n \phi, \cos n \phi,0) 
\ , \label{rtp} \end{eqnarray}
respectively.
 
For $n=1$, $H_1=H_3=0$, $H_2=H_4=a(r)$, $f=f(r)$ and $g=\theta$
the spherically symmetric ansatz of \re{spha}-\re{sphf} is recovered.

The residual $U(1)$ gauge degree of freedom \cite{KKT} 
is fixed by the condition $r\partial_r H_1-\partial_\theta H_2 =0$,
which is just the Coulomb gauge in the two dimensional residual $U(1)$
subsystem resulting from the imposition of radial symmetry in the
$x-y$ plane, i.e. with radius $\rho=\sqrt{x^2+y^2}$.

At the origin the boundary conditions for the gauge field functions read
\begin{equation}
H_2|_{r=0}=H_4|_{r=0}= 1, \ \ \ H_1|_{r=0}=H_3|_{r=0}=0
\ , \label{bc2c} \end{equation}
and for the Skyrme functions
\begin{equation}
f|_{r=0} = \pi, \ \ \ \partial_r g|_{r=0}=0 
\ . \label{bc2b} \end{equation}

For the gauge field to approach the asymptotic configuration of a monopole
we choose
\begin{equation}
H_2|_{r=\infty}=H_4|_{r=\infty}=0, \ \ \ 
H_1|_{r=\infty}=H_3|_{r=\infty}=0
\ , \label{bc1c} \end{equation}
and for the Skyrme field functions 
\begin{equation}
f|_{r=\infty}=\omega, \ \ \  g|_{r=\infty}=\theta
\ . \label{bc1b} \end{equation}

The boundary conditions along the $\rho$- and $z$-axis
are determined by the symmetries.
For the gauge field functions symmetry considerations
lead to the boundary conditions
\begin{equation}
\begin{array}{lllllllllll}
H_1|_{\theta=0}&=&H_3|_{\theta=0}&=&0 &\ , \ \ \ &
\partial_\theta H_2|_{\theta=0} &=& \partial_\theta H_4|_{\theta=0} 
 &=& 0 \ ,
\\
H_1|_{\theta=\pi/2}&=&H_3|_{\theta=\pi/2}&=&0 &\ , \ \ \ &
\partial_\theta H_2|_{\theta=\pi/2} &=& 
\partial_\theta H_4|_{\theta=\pi/2} &=& 0
\end{array}
\   \label{bc4c} \end{equation}
along the axes,
as well as the condition 
\begin{equation}
H_2|_{\theta=0}=H_4|_{\theta=0}
\ . \label{h2h4} \end{equation}
Along these axes the Skyrme field functions satisfy the following boundary
conditions;
\begin{equation}
\begin{array}{lllll}
\partial_\theta f|_{\theta=0} = 0 \ , \ \ \
 \partial_\theta g|_{\theta=0} = 1
\ , \\
\partial_\theta f|_{\theta=\pi/2} = 0 \ , \ \ \
\partial_\theta g|_{\theta=\pi/2} = 1  
\end{array}
\   \label{bc4b} 
\end{equation}

\subsection{Numerical results}
Solving the set of partial differential equations numerically for
the model characterised by $\omega=\pi/2$, and for different values
of $\kappa$, we find that even for
$\kappa=0$ there is only an attractive phase. As shown in Fig. 4, the
difference $\delta E$ of the energy per winding number $n$ between the 
$n=1$ and $n=2$ increases with increasing $\kappa$. This is an
indication that in this model 2-monopole bound states can exist. Although
a definitive demonstration of this is well beyond the scope of the
present work. (That would involve finding the dependence of the
interaction energy on the separation of the two monopoles.)

Moreover, we calculated the energy $E_{n}(\kappa)$, for different values
of $\kappa$ and $n$. The values are given in the tables below.
Unfortunately the numerical process becomes less reliable with
increasing monopole
charge $n$. As a result we restrict our numerical constructions to
$n\le 4$ only. For $\kappa=0$, $\kappa=3$ and $\kappa=5$ we find
\begin{center}
\begin{tabular}{|l|c|c|c|c|}
\hline \hline
$n$ & $1$ & $2$ & $3$ & $4$ \\
\hline
$E_n(0)$ & 2.95 & 4.82 & 6.63 & 7.56  \\
\hline \hline
\end{tabular}
\begin{center}

\begin{tabular}{|l|c|c|c|c|}
\hline \hline
$n$ & $1$ & $2$ & $3$ & $4$ \\
\hline
$E_n(3)$ & $3.40$ & $5.40$ & $7.30$ & $8.80$ \\
\hline \hline
\end{tabular}
\medskip

\begin{tabular}{|l|c|c|c|c|}
\hline \hline
$n$ & $1$ & $2$ & $3$ & $4$ \\
\hline
$E_n(5)$ & $3.48$ & $5.60$ & $7.50$ & $9.31$ \\
\hline \hline
\end{tabular}
\end{center}
\end{center}

>From this data we can deduce some quantitative information on the
interaction energies of the monopoles leading to the possible formation
of lumps of charges $n\le 4$. To this end we define
the following 'binding energy' corresponding to the
energy needed to dissociate a charge-$n$ lump into a charge-$n-1$ and a
charge-$1$ lump, divided by the energy of the charge-$n$ lump.
\be
\Delta E^{\{n-1,1\}}_n(\kappa)=\frac{[E_{n-1}(\kappa)
+E_1(\kappa)]-E_n(\kappa)}{E_n(\kappa)}    \ .
\ee
The values for different $\kappa$ and $n$ are given in the
Table below.\\
\begin{center}
\begin{tabular}{|l|c|c|c|}
\hline \hline
$ n $ & $\Delta E^{\{n-1,1\}}_n(0)$ & $\Delta E^{\{n-1,1\}}_n(3)$ &
$\Delta E^{\{n-1,1\}}_n(5)$ \\
\hline
2 & 0.22 & 0.26 & 0.24 \\
3 & 0.17 & 0.21 & 0.21 \\
4 & 0.27 & 0.22 & 0.18 \\
\hline \hline 
\end{tabular}
\end{center}
This table shows that for all three values of $\kappa$,
$\Delta E^{\{n-1,1\}}_n(\kappa)$
remains positive, which is an indication for the existence of
monopole lumps of charges up to $n=4$.

It is hard to extract any reliable conclusion from such meagre data,
but the fact that the binding energies do not seem to decrease with
increasing $n$ is encouraging from the point of view of the possibility
of finding very large monopole clumps. This question will be investigated
elsewhere, using different numerical techniques.

\section{Summary and Discussion}

We have studied a particular variant of the Skyrme model gauged
according to the prescription \re{cov}, equivalent to the commutator
gauging with respect to the $SU(2)$ gauge connection
$A_i=-\frac{i}{2}\vec A_i\cdot\vec \tau$,
\be
\label{covu}
D_iU=\pa_iU+[A_i,U]\ ,
\ee
which is augmented by the potential \re{pot}. The function of this
potential is to fix the boundary value of the Skyrme field at large
distances which, unlike in the usual (ungauged) Skyrme model, is not
fixed by the requirement of finite energy\footnote{Unlike here, in
Refs.\cite{AT,BT}, an explicit potential was not employed. This is
because there, {\em only one} value, $\omega=0$, was used so it was
considered reasonable to treat the role of the corresponding potential
in a putative capacity. This is justified since, as we have verified here
too, the explicit use of a potential only affects the asyptotics of the
chiral function $f(r)$ and otherwise leads to no qualitative
differences.}. Thus we have considered a
set of the gauged Skyrme models characterised by the parameter $\omega$
appearing in the potential \re{pot}. In the usual Skyrme model $\omega=0$.

The qualitative properties of the solitons, depending on the parameters
$\omega$ and $\kappa$ characterising the models, are studied in the
spherically symmetric case.

The main effect of the boundary condition $\omega\neq 0$ is the breaking
of the $SO(3)$ gauge symmetry of the solution down to $SO(2)$,
asymptotically. This is related to the nonvanishing VEV of the Skyrme
field $\phi^{\alpha}$. This
can be seen clearly from the second member of \re{chirasym}. In addition
to this magnetic charge, a noninteger version of the Baryon number
given by \re{bar} is also associated with this solution. We have verified
the existence of finite energy solutions bounded from below by both these
charges for the allowed ranges of $\omega$. The latter (ranges) depend
also on the (effective) Skyrme coupling $\kappa$ of the model. These
ranges have been illustrated in Fig. 1.

A surprising if not unexpected property of these models is, that the
model characterised by $\kappa=0$ does support a soliton. This could
have been expected since in the presence of the Yang--Mills term, it
is not necessary to have a Skyrme term to satisfy the (Derrick) scaling
requirement. We found that the solitons of the $\kappa=0$ models are
generally quite similar to those of the the models with $\kappa\neq0$,
with one noticable qualitative difference. This concerns the restricted
range of allowed $\omega$ for the $\kappa=0$ model as seen in Fig. 1,
(and in Fig. 3)
which in addition to this qualitative feature also illustrates the fact
that for small enough $\kappa$ the profile of the chiral function $f(r)$
sinks below its asymptote $\omega$ and approaches it from below.

By contrast when $\omega=0$, the model with $\kappa=0$ cannot support a
soliton because in that case the lower bound \re{bogo} disappears,
leaving only the lower bound \re{bar} in place, and the latter trivialises
in the $\kappa=0$ limit. This is also borne out by the graphs in Fig. 1.

After exposing the qualitative features of the solitons in the spherically
symmetric case, we studied axially symmetric solutions of the model. Our
aim here was to discover if like monopoles of the $\omega\neq0$ models
are in attractive or repulsive phases. For this purpose we restricted
ourselves to the $\omega=\frac{\pi}{2}$ model which has the nice feature
of having integer magnetic charge. We did however consider varying values
of the (effective) Skyrme parameter $\kappa$, including the distinguished
case of $\kappa=0$.

The main interest in the model specified by $\kappa=0$ in this respect is
that, like the usual (ungauged) Skyrme model, it exhibits no free coupling
constants that can parametrise the crossover from an attractive to a
repulsive phase. Like the latter it lacks also a neutral, or Bogomol'nyi
saturated phase, where the (static Euclidean) stress--energy tensor would
have vanished leading to noninteracting solitons. Thus the unique phase in
which the solitons are supported is of special interest, especially if
it were attractive because then the switching on of $\kappa$ would most
likely not result in a crossover to a repulsive phase.
We have verified that this is precisely what happens, as illustrated
in Fig. 4 for the magnetic charge-2 soliton. An
outstanding problem in this context is the calculation of the (attractive)
interaction energy of two 1-monopoles as a function of their separation.
This would demonstrate the existence of bound states definitively.

In addition to verifying that these models support mutually attracting
like\\
monopoles, we sought some indications as to whether there is the
possibility of forming bound states of monopole charge greater than 2.
(Apart from its intrinsic interest, the formation of very large monopole
clumps may be relevant in cosmology.) To this end, axially symmetric
solutions with monopole charges $n=2,3,4$, for the $\kappa=0$,
$\kappa=3$, and the $\kappa=5$ models were constructed. It was seen that
all these solitons remained in attractive phases. While it is expected
that these axially
symmetric solitons with $n>2$ are not the lowest energy solutions, the
latter probably exhibiting (solid) Platonic symmetries~\cite{BTC,BS} as
in the usual Skyrme model, it is nonetheless true that they give a
reliable indication towards the existence of such bound states.
To this end we list the binding energies against the dissociation of
an axially symmetric charge--$n$ monopole into a charge--$n-1$ and a
charge--$1$ monopole in the last Table of Section {\bf 4}. We see that
this binding energy stays positive and does not change too much as $n$
increases, at least for $n\le 4$. Unfortunately the numerical process was
not reliable much beyond $n=4$.

While in the present work our primary aim has been the qualitative
study of the solitons of the gauged Skyrme model charcterised by
$\omega\neq0$, and the resulting property of the interaction of like
monopoles in these theories, it may be in order to emphasise some
physically attractive features of these models. These models combine
some attractive features of (a) Higgs models,
in this case the Georgi--Glashow model in that they support monopoles,
and features of (b) Skyrme models, in this case the usual Skyrme
model~\cite{S} in that the like-charged solitons are in an attractive
phase. Indeed it appears that the attraction properties of these models
are considerably more pronounced than those of the Skyrme model~\cite{S}.

Apart from the practical consideration of the possibility of supporting
large\\
monopole clumps, there are two theoretical properties of the
models that deserve mention. One is the physically desirable
property of the symmetry breaking from $SU(2)$ to $U(1)$,
and the other one is the fact that the $\kappa=0$ model supports
solitons with much the same qualitative properties as the generic models.
The considerable advantage of this is that we avail of the Skyrme
theortic feature of mutually attracting solitons without having to pay
the price of featuring a (quartic kinetic) Skyrme term in the Lagrangian,
thus avoiding the attendant severe problems of quantisation.

\medskip

\noindent
{\bf Acknowledgements:} We are indepted to B. Kleihaus for his
collaboration at the initial stage of this investigation. This work
was carried out in the framework of projects SC/96/602 and IC/00/021
of Enterprise--Ireland.
We would like to thank the RRZN Hannover for computing time.

\newpage

\small{
 }
\newpage

\begin{figure}
\centering
\epsfysize=11cm
\mbox{\epsffile{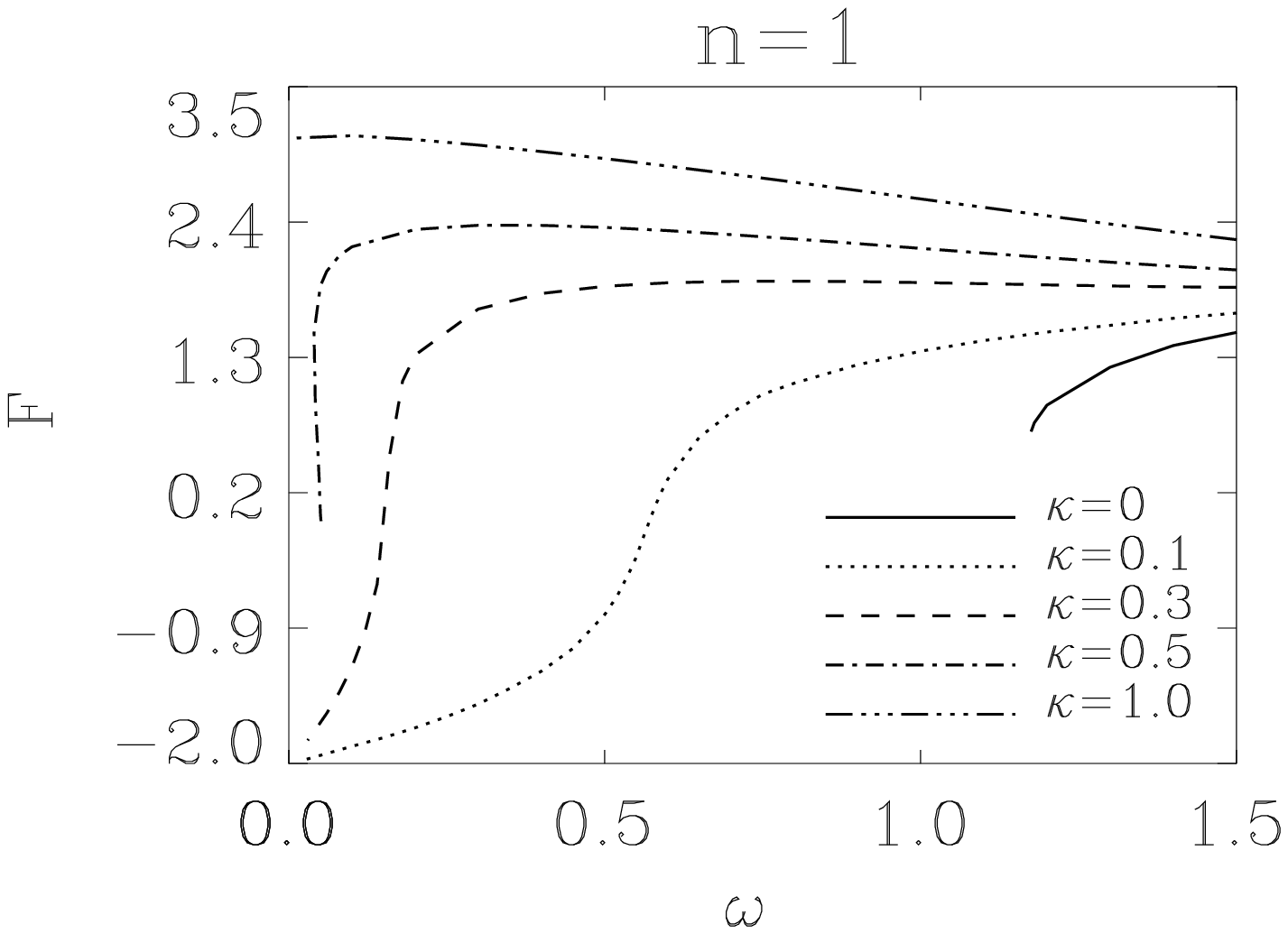}}
\caption{ The quantity $F$, that determines the
asymptotic behaviour of the skyrme field function $f$
with $f(x\gg 1)=\omega+F/x+o(1/x^{2})$ is shown as a
function of $\omega$ for $n=1$ and different values of $\kappa$.}
\end{figure}
\newpage

\begin{figure}
\centering
\epsfysize=11cm
\mbox{\epsffile{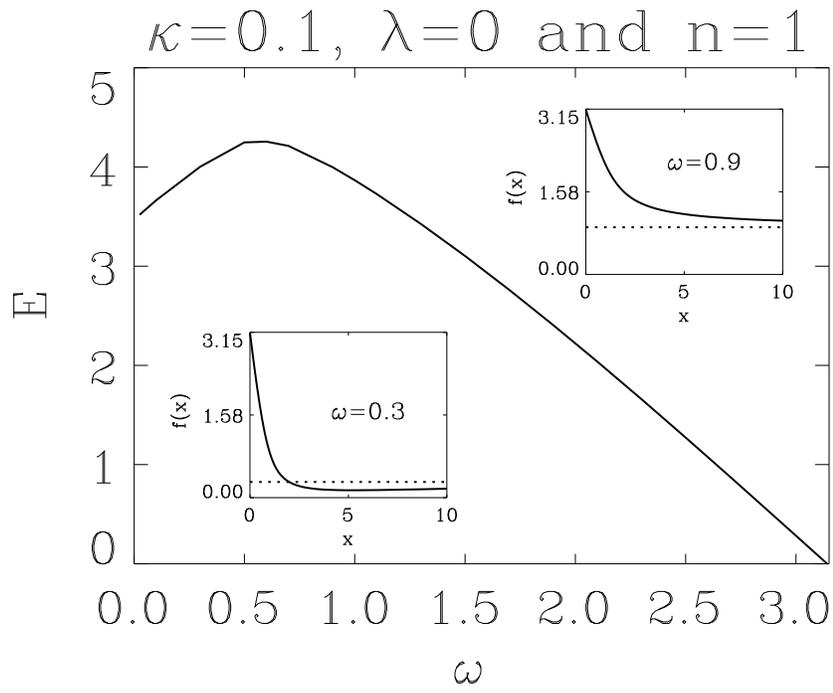}}
\caption{
The energy $E$ of the $n=1$ solution is shown as a function of 
$\omega$ for $\kappa=0.1$ and $\lambda=0$.
The two inlets show the skyrme field function $f(x)$ as function of
the radial coordinate $x$
for a) $\omega=0.3 < \omega_{max}$  and  b) $\omega=0.9 >
\omega_{max}$.
}
\end{figure}
\newpage

\begin{figure}
\centering
\epsfysize=11cm
\mbox{\epsffile{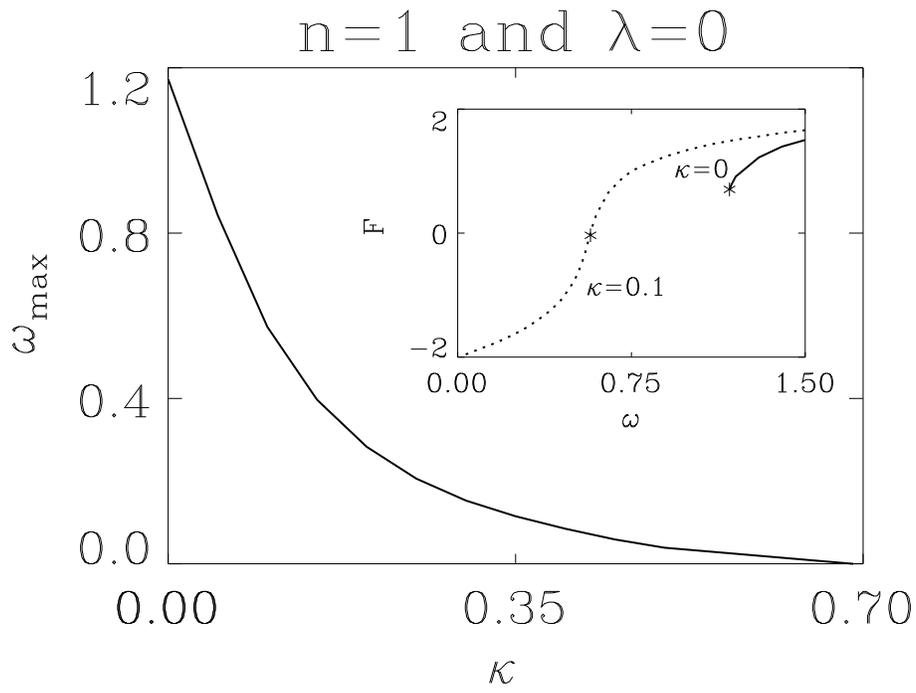}}
\caption{$\omega_{max}$, the value of $\omega$ at which the energy has its maximum, 
is shown as function of $\kappa$.
The inlet shows the quantitiy $F$ (see Fig.$1$) over $\omega$ for $\kappa=0$ and
$0.1$. The asteriks mark $\omega_{max}(\kappa)$.}
\end{figure}
\newpage

\begin{figure}
\centering
\epsfysize=11cm
\mbox{\epsffile{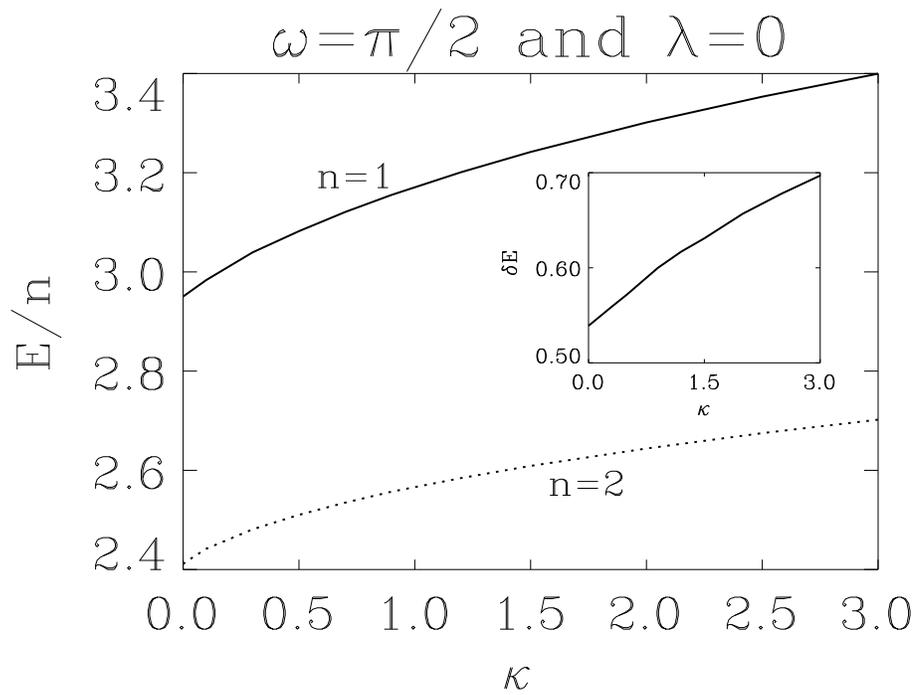}}
\caption{
The energy per winding number $E/n$
is shown as a function of $\kappa$
for $n=1$ and $n=2$ , $\omega=\pi/2$ and $\lambda=0$.
The inlet shows the difference
of the energy per winding number $\delta E=E(n=1)-E(n=2)/2$
between the $n=1$ and the $n=2$
solution.
}
\end{figure}
\end{document}